\documentclass{nature}
\usepackage{amsmath,amssymb,graphicx,hyperref,url}  
\usepackage[none]{hyphenat}
\usepackage[usenames]{color}

\newcommand{\beq}{\begin{equation}}
\newcommand{\eeq}{\end{equation}}
\newcommand{\ben}{\begin{eqnarray}}
\newcommand{\een}{\end{eqnarray}}
\newcommand{\bi}{\begin{itemize}}
\newcommand{\ei}{\end{itemize}}
\newcommand{\pdf}{p}
\newcommand{\data}{{d}}

\newcommand{\mdl}{{M}}

\newcommand\ee{\end{equation}}
\newcommand\be{\begin{equation}}
\newcommand\eea{\end{eqnarray}}
\newcommand\bea{\begin{eqnarray}}

\title{Planck evidence for a closed Universe and a possible crisis for cosmology}

\author{Eleonora Di Valentino$^1$,
Alessandro Melchiorri$^2$, Joseph Silk$^{3,4,5,6}$}

\begin{document}

\maketitle

\begin{abstract}
The recent Planck Legacy 2018 release has  confirmed the presence of an enhanced  lensing amplitude in CMB power spectra compared to that predicted in the standard $\Lambda$CDM model. A closed universe can provide a physical explanation for this effect, with the Planck CMB spectra now preferring a positive curvature at more than $99 \%$ C.L. Here we further investigate the evidence for a closed universe from Planck, showing that positive curvature naturally explains the anomalous lensing amplitude and demonstrating that it also removes a well-known tension within the Planck data set concerning the values of cosmological parameters derived at different angular scales. We show that since the Planck power spectra prefer a closed universe, discordances higher than generally estimated arise for most of the local cosmological observables, including BAO. The assumption of a flat universe could therefore mask a cosmological crisis where disparate observed properties of the Universe  appear to be mutually inconsistent. Future measurements are needed to clarify whether  the observed discordances are due to undetected systematics, or to new physics, or simply are a statistical fluctuation.

\end{abstract}


The recent Planck Legacy 2018 release of observations of Cosmic Microwave Background (CMB) anisotropies (PL18, hereafter) has reported some unexpected results, revealing  the possibility for new physics beyond the so-called $\Lambda$CDM standard cosmological model \cite{planck2018,planck20182}. Indeed, while the inflationary predictions for coherent acoustic oscillations have been fully confirmed, a preference for a higher lensing amplitude $A_{lens}$ than predicted in the base $\Lambda$CDM at about $3$ standard deviations has been found in the temperature and polarization angular spectra. We argue that the "$A_{lens}$" anomaly has profound implications for some extensions to $\Lambda$CDM such as the curvature of the universe. The constraints from the PL18 CMB spectra on curvature, parameterized through the energy density parameter $\Omega_K$, are indeed quite surprising, suggesting a closed universe at $3.4$ standard deviations ($-0.007>\Omega_K>-0.095$ at $99 \%$ C.L.~\cite{planck2018,planck20182,planckwiki}).  

As is well known, inflation theory naturally predicts a flat universe~\cite{linde1981,as1981}. However, inflationary models with $\Omega_K<0$ ~\cite{linde1995,omegainf1,uzan} are relatively simple to build, with primordial homogeneity and isotropy easier to achieve than in open models. An issue for closed inflation models is that to obtain $\Omega_K\sim-0.1$, fine-tuning at a few per cent level is needed~\cite{omegainf1}. This does not sound very compelling, but it may still be acceptable given the presence of a far more 
finely-tuned cosmological constant. Closed models could also lead to a large-scale cut-off in the  primordial density fluctuations, around the curvature scale $R_c=(c/H_0)|\Omega_K|^{-0.5} \sim 10 \rm \, Gpc$, in agreement with the observed low CMB anisotropy quadrupole~\cite{omegainf1,Efstathiou:2003hk}. Confirmation of a positive spatial curvature would also have several implications for inflationary theory, and, for example, severely challenge  models of eternal inflation~\cite{2006JHEP...03..039F, 2012PhRvD..86b3534G}.

In this Letter, we show that, if indeed credible, the Planck preference for a closed universe introduces a new problem for modern cosmology. 
Indeed, many of the current tight constraints on cosmological parameters are obtained by combining complementary data sets. A basic assumption in this procedure is that these different data sets must be consistent, i.e., they must plausibly arise from the same cosmological model. Currently, there are two major experimental data sets that are  in tension with Planck: the determination of the Hubble constant by Riess et al. 2018~\cite{riess2018}  is discrepant at the level of $\sim 3$ standard deviations (but see also \cite{riess2019}), and the observations of cosmic shear by the KiDS-450 survey  disagree at about two standard deviations~\cite{Hildebrandt:2016iqg,joudaki16}.
 Furthermore, the value of $A_{lens}$ derived from the 
Planck lensing-generated 4-point correlation function is consistent with the expectations of $\Lambda$CDM and in tension with the PL18 power spectra~\cite{planck2018,Motloch:2018pjy}. 

While most of the remaining cosmological observables are considered to be  in good agreement with PL18, these inconsistencies have already motivated several studies that  attempt to critically reassess the level of discordance~\cite{tool1,tool2,tool3},  or to resolve it with the introduction of new physics~\cite{newphys1,newphys2,newphys4,newphys7,Yang:2018qmz}.
 
The level of accordance between cosmological observables has hitherto been thoroughly investigated under the assumption of a flat universe. We show here that  when curvature is allowed to vary (as suggested by the PL18 CMB spectra), the statistical significance of the known tensions with PL18 increases, and in addition, other discrepancies arise with several "local" (i.e., at redshift $z<3$) observables. The assumption of a flat universe could, therefore, mask a cosmological crisis where disparate observed properties of the Universe  appear to be  mutually inconsistent.

\begin{figure*}
\begin{center}
\includegraphics[width=0.57\linewidth]{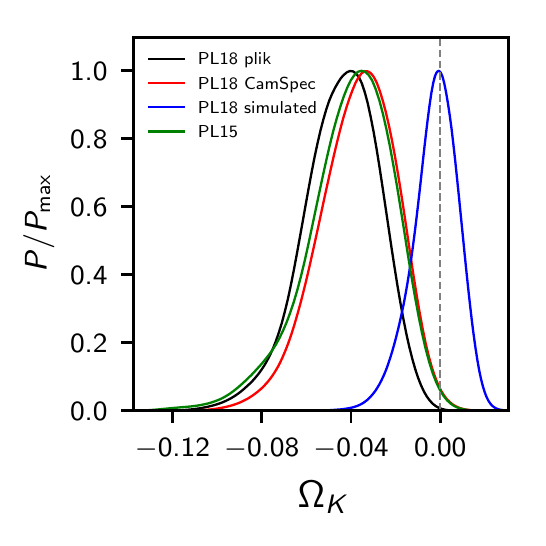}
\end{center}
\caption{{\bf Preference for a closed universe, $\Omega_k<0$, from Planck.} Posterior distributions on the curvature density parameter $\Omega_K$ from Planck 2018 (PL18) temperature and polarization-simulated angular power spectra (assuming a fiducial flat $\Lambda$CDM model) and PL18 real data adopting the baseline Planck likelihood and the alternative CAMSPEC likelihood, respectively. For comparison, the posterior from the previous Planck 2015~\cite{planck2015} (PL15) data release is also shown.}
\label{figure1}
\end{figure*} 

Before evaluating the tensions of the PL18 results with independent cosmological observables, we first check whether the PL18 power spectra can provide an unbiased and reliable estimate of the curvature of the Universe. This may not be the case since  "geometrical degeneracy" is present between cosmological parameters~\cite{Bond:1997wr,Efstathiou:1998xx,Elgaroy:2007bv}. For example, assuming the same inflationary parameters and reionization process, a flat cosmological model with matter density $\Omega_m = 0.35$, cosmological constant density $\Omega_{\Lambda} = 0.65$, and Hubble constant $H_0=65\, \rm km/s/Mpc$ produces an identical structure of the CMB angular spectrum at sub-degree angular scales for a closed model with $\Omega_m= 1$, $\Omega_{\Lambda}=0.15$ (i.e. $\Omega_K=-0.15$) and $H_0=38.4\, \rm km/s/Mpc$. Because of the form of the degeneracy, different closed models have identical CMB power spectra to that of  a single flat model.  The main consequence is that, after marginalization over the nuisance parameters, the posterior on $\Omega_K$ is generally skewed towards closed models~\cite{melgri,wmap9}.

The situation changes with precise CMB measurements at arc-minute angular scales: here, indeed, additional anisotropies induced by gravitational lensing are not negligible. Since gravitational lensing depends on the matter density, its detection breaks the geometrical degeneracy. The Planck experiment with its improved angular resolution therefore offers the opportunity of a precise measurement of curvature from a single CMB experiment.

To confirm this hypothesis, we generated a Monte Carlo Markov Chain analysis over simulated Planck (temperature and polarization) data, assuming the best-fit flat $\Lambda$CDM model and experimental noise properties similar to those presented in the PL18 release~\cite{planck2018}. As we can see from Figure~\ref{figure1}, the expected posterior is centered around $\Omega_K=0$ with a bound of $\Omega_K=0.00\pm0.02$ at $68 \%$ C.L.. Potentially, an experiment such as Planck could constrain curvature with  $\sim 2\%$ uncertainty, without any significant bias towards closed models.

For comparison, in Figure~\ref{figure1} we have plotted the posterior from the PL18 real temperature and polarization power spectra, assuming the baseline Planck likelihood (see \cite{planck20182}). As we can see, the posterior is reasonably centered on a closed model around $\Omega_K=-0.04$. Integrating this posterior distribution over $\Omega_K$, we find that Planck favors a closed Universe ($\Omega_K<0$) with  $99.985 \%$  probability. Moreover, a closed universe with $\Omega_K=-0.0438$ provides a better fit to PL18 with respect to a flat model, with a $\chi^2$ difference of $\Delta \chi_{eff}^2 \sim -11$~\cite{planckwiki}. 

This qualitatively shows the PL18 preference for a closed Universe, but  does not  statistically weight the additional parameter 
($\Omega_K$). To better quantify the preference for a closed model, we adopt the Deviance Information Criterion~\cite{dic1,dic2,dic3} (DIC)  that takes into account the Bayesian complexity, i.e., the effective number 
of parameters, of the extended model~\cite{dic2}, and is  defined as:

\begin{equation}
    DIC=2\overline{\chi^2_{eff}}-\chi^2_{eff}
\end{equation}

\noindent  where $\chi^2_{eff}$ is the best-fit chi-square from the MCMC chains and 
the bar denotes a mean over the posterior distribution. This latter quantity can be easily computed. 
We restrict the analysis to models with curvature in the range $-0.2\le\Omega_K\le0$, i.e.
we neglect open models since they are both disfavored from observations and more difficult to realize in
an inflationary scenario. We find that the Planck data yields $\Delta DIC=-7.4$, i.e. a closed universe 
with $\Omega_K=-0.0438$ is preferred with a probability ratio of about $1:41$ with respect to a flat model.

We also compute the Bayesian Evidence ratio by making use of the Savage-Dickey density ratio 
(SDDR)~\cite{dic2,SDDR,Trotta2}.
Assuming SDDR the Bayes factor $B_{01}$ can be written as 

 \begin{equation} \label{eq:savagedickey}
 B_{01} = \left.\frac{\pdf(\Omega_K \vert \data, \mdl_1)}{\pi(\Omega_K |
 \mdl_1)}\right|_{\Omega_K = 0} \quad {\text{(SDDR)}}.
 \end{equation}
\noindent where $\mdl_1$ denoted the model with curvature, 
$\pdf(\Omega_K \vert \data, \mdl_1)$ is the posterior for $\Omega_K$ in this theoretical
framework computed from a specific data set $\data$, and 
$\pi(\Omega_K | \mdl_1)$ is the prior on $\Omega_K$ that we assume as flat in the
range $-0.2\le\Omega_K\le0$.

Applying the Savage-Dickey method to the Planck temperature and polarization, we obtain the Bayes ratio of:

\begin{equation}
    |\ln B_{01}| = 3.3
\end{equation}

\noindent i.e. we obtain, assuming the so--called "Jeffrey's scale",  strong evidence for 
closed models with $\Omega_K$ in the prior range $[-0.2,0]$. While the assumption of a larger prior would lead to  weaker evidence, the preference from the data   for a closed universe is clear.

This evidence could come from an unidentified systematic in the Planck data. However, as we can also see from the
posteriors in Figure~\ref{figure1}, the preference for a closed Universe increases as we  move from the Planck 2015\cite{planck2015} (PL15) to the current PL18 release. Moreover, even assuming a significantly different procedure for the likelihood analysis~\cite{planck20182}, and using the alternative CAMSPEC approach instead of the baseline Planck likelihood, the preference for curvature is reduced but is  still well above two standard deviations with $\Omega_K=-0.037_{-0.034}^{+0.032}$ at $95 \%$ C.L.~\cite{planckwiki}. We find, in  the case of CAMSPEC, $\Omega_K<0$ with a $99.85 \%$ probability. While the indication for a closed universe is less significant with CAMSPEC, it is still present, showing that our result  is not due to differences between analysis methods.

\begin{figure*}[!hbtp]
\begin{center}
\includegraphics[width=.4\textwidth,keepaspectratio]{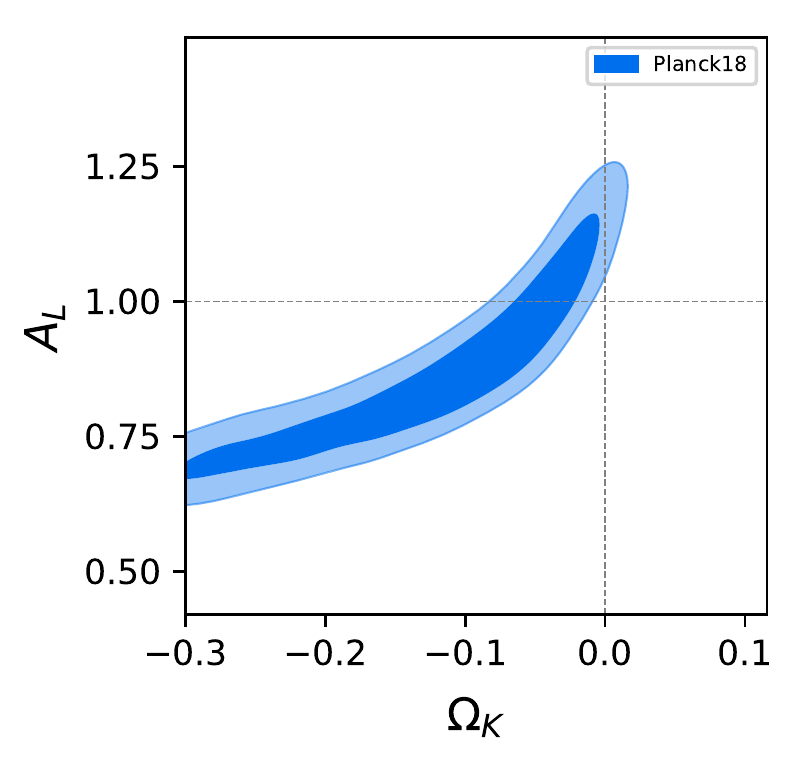}
\end{center}
\caption{{\bf Degeneracy between curvature and lensing}. Constraints at $68 \%$ and $95 \%$ in the $A_{lens}$ vs $\Omega_K$ plane from Planck 2018 temperature and polarization data. A degeneracy between curvature and the $A_{lens}$ parameter is clearly present. Note that a model with $\Omega_K<0$ is slightly preferred with respect to a flat model with $A_{lens}>1$.}
\label{figure2bb}
\end{figure*}

\begin{figure*}[!hbtp]
\begin{center}
\includegraphics[width=.70\textwidth,keepaspectratio]{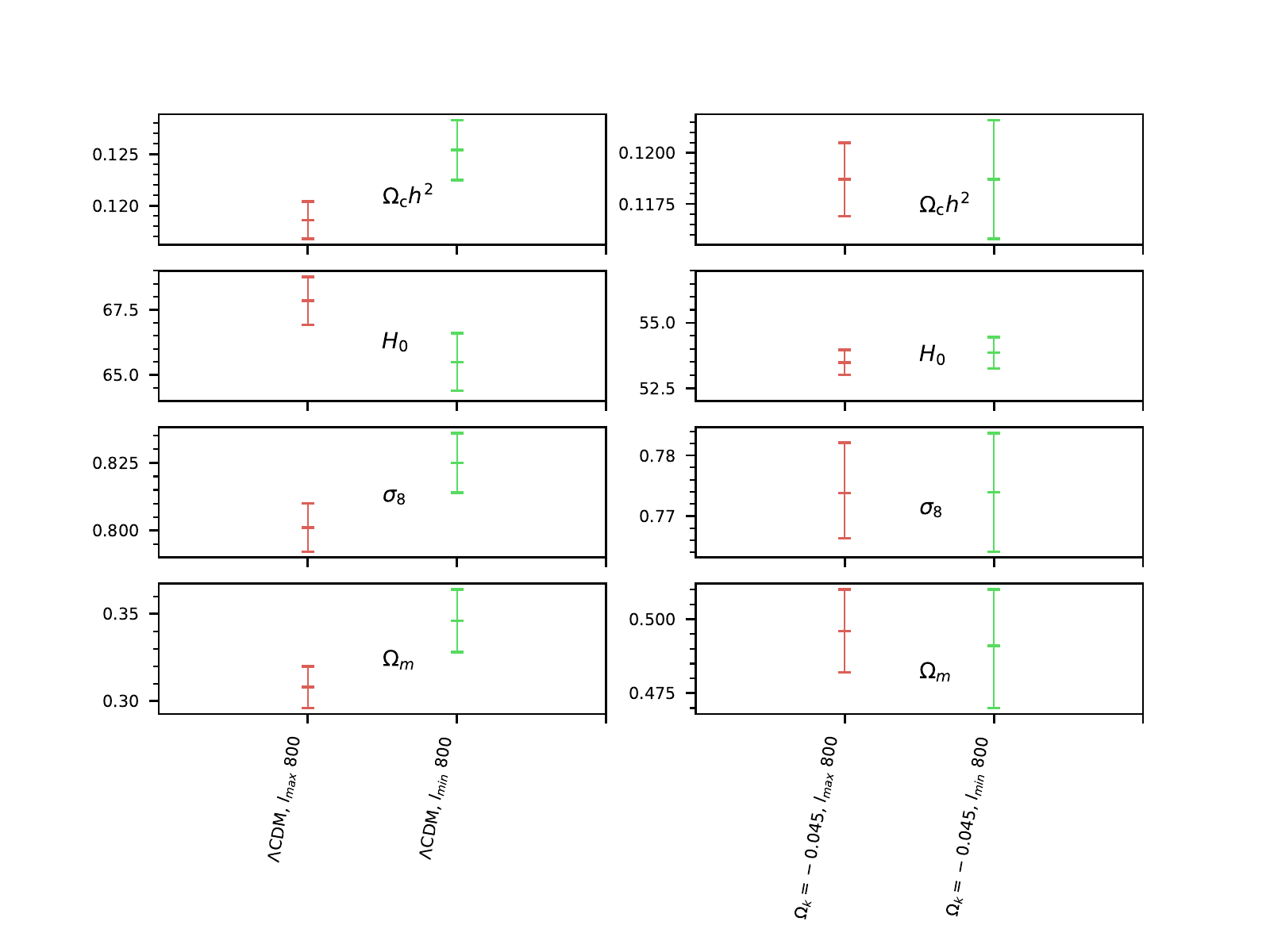}
\end{center}
\caption{{\bf Curvature and parameters shift}. Cosmological parameters derived from two different multipoles ranges ($2\le\ell\le800$ and
$800<\ell\le2500$) of the Planck 2018 temperature and polarization data assuming either a $\Lambda$CDM model (Left) or  a closed model (Right). Polarization data at low multipoles ($2\le \ell \le 30$) is included in both cases.
The difference in the parameter constraints present in flat $\Lambda$CDM, disappears when assuming a model with $\Omega_k=-0.045$.}
\label{figure2bc}
\end{figure*}

The preference for closure in the Planck data is strongly connected with the higher lensing amplitude. This is evident from the parameter degeneracy between $A_{lens}$ and $\Omega_K$ as shown in Figure~\ref{figure2bb} where we report the 2D constraints at $68\%$ and $95 \%$ C.L. on $A_{lens}$ and $\Omega_K$ from the Planck 2018 temperature and polarization data\cite{planck20182}.
The dark matter content can indeed be greater in a closed universe, leading to a larger lensing signal, solving the $A_{lens}$ anomaly, and providing a robust physical explanation. As we can see, when a closed model is considered, $A_{lens}$ is in agreement with the expectation of $A_{lens}=1$. The amplitude of the lensing signal in Planck temperature and polarization data is precisely what is expected in a closed universe. It is interesting to note that a $\Lambda$CDM+$\Omega_K$ analysis provides a marginally better fit to the $\Lambda$CDM+$A_{lens}$ analysis by $\Delta \chi^2=-1.6$, due to the fact that closed models better fit the low multipole data.

As discussed in \cite{addison}, assuming flat $\Lambda$CDM, the values of the cosmological parameters obtained from the
Planck 2015 temperature angular spectrum in the multipole range $\ell<800$ are "shifted" with respect to those derived from
the same Planck data relative to multipoles in the range $800<\ell<2500$. This tension is also present in the PL18 release\cite{planck20182} and the inclusion of the $A_{lens}$ parameter removes this difference. A key point of our paper is that  the addition of curvature also solves this tension: in Figure~\ref{figure2bc} we show that in a closed universe with $\Omega_K=-0.045$ the cosmological parameters derived in the two different multipole ranges,  from Planck 2018 temperature and polarization data are now fully compatible.

\begin{figure*}[!hbtp]
\begin{center}
\includegraphics[width=.70\textwidth,keepaspectratio]{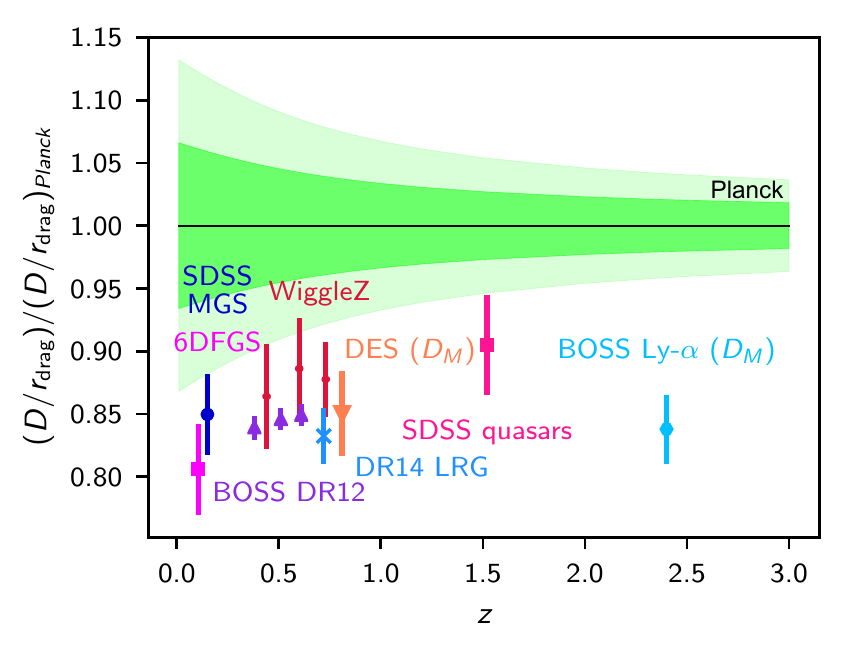}
\end{center}
\caption{{\bf Tension with BAO}. Acoustic-scale distance measurements divided by the corresponding mean distance ratio from PL18 temperature and polarization power spectra in a $\Lambda$CDM$+\Omega_K$ model. The green bands show the $68\%$ and $95\%$ confidence ranges. The data points correspond to  the measurements at $68\%$ C.L. from the following experiments: 6dFGS~\cite{6dFGS}, SDSS MGS~\cite{mgs}, and BOSS DR12~\cite{bossdr12} (the BAO dataset considered in this paper). We also report measurements from WiggleZ~\cite{wigglez}, DES~\cite{des}, DR14 LRG~\cite{lrg}, SDSS quasars~\cite{sdssqso}, and BOSS Lyman-$\alpha$~\cite{bosslya}.}
\label{figure2}
\end{figure*}

However, if the PL18 power spectra suggest a closed universe, the remaining cosmological observables are in strong disagreement with this. Let us now compare the Planck constraints with those coming from local observables, starting with baryon acoustic oscillations.
We first consider a combination of measurements given by the 6dFGS~\cite{6dFGS}, SDSS-MGS~\cite{mgs}, and 
BOSS DR12~\cite{bossdr12} surveys (hereafter we refer to this dataset simply as BAO), as 
adopted by the Planck collaboration~\cite{planck2018}. The combination of this BAO dataset and PL18 power spectra produces a  strong constraint on curvature with $\Omega_K=0.0008_{-0.0037}^{+0.0038}$ at $95 \%$ C.L.~\cite{planck2018}, in excellent agreement with a flat Universe. Given the significant change in the conclusions from Planck alone, it is reasonable to investigate whether  the BAO dataset is actually consistent with PL18. The level of concordance between Planck and BAO, even from a qualitative point of view, is immediately clear from Figure~\ref{figure2} where we plot the acoustic-scale distance ratio $D_V(z)/r_{drag}$ as a  function of redshift $z$, taken from several recent BAO surveys, and divided by the mean acoustic-scale ratio obtained by Planck temperature and polarization data adopting a $\Lambda$CDM$+\Omega_K$ model. We note that $r_{drag}$ is the comoving size of the sound horizon at the time of the end of the baryon drag epoch and $D_V$ is a combination of the Hubble parameter $H(z)$ and the comoving angular diameter distance $D_M(z)$: $D_V(z)=(czD_M^2(z)/H(z))^{1/3}$. As we  see, there is a striking disagreement between the PL18 power spectra and BAO. This can also be seen in Table~\ref{Table1} where we report the constraints on $D_M$ and $H(z)$ from the recent analysis of BOSS DR12 data~\cite{bossdr12} and the corresponding constraints obtained indirectly from Planck assuming a $\Lambda$CDM model with curvature. Each of the BOSS DR12 data points is in disagreement by about $\sim 3$ standard deviations with the Planck power spectra.

\begin{table*}
\begin{center}\footnotesize
\begin{tabular}{|l|c|c|c|c|}
\hline
Observable & Redshift& BAO ($68 \%$ C.L.)&Planck ($68 \%$ C.L.)& Tension\\
\hline
$D_M (r_{d,fid}/r_d)$ [Mpc] &$z=0.38$&$1518\pm22.8$&$1843\pm100$&$2.9\sigma$\\
\hline
$D_M (r_{d,fid}/r_d)$ [Mpc] &$z=0.51$&$1977\pm26.9$&$2361\pm115$&$3.0\sigma$\\
\hline
$D_M (r_{d,fid}/r_d)$ [Mpc] &$z=0.61$&$2283\pm32.3$&$2726\pm130$&$3.3\sigma$\\
\hline
$H (r_{d,fid}/r_d)$ [km/s/Mpc] &$z=0.38$&$81.5\pm1.9$&$71.6\pm3.3$&$2.6\sigma$\\
\hline
$H (r_{d,fid}/r_d)$ [km/s/Mpc] &$z=0.51$&$90.5\pm1.97$&$78.9\pm3.1$&$3.1\sigma$\\
\hline
$H (r_{d,fid}/r_d)$ [km/s/Mpc] &$z=0.61$&$97.3\pm2.1$&$85.0\pm3.0$&$3.3\sigma$\\
\hline 
\end{tabular}
\end{center}
\caption{\bf Comoving angular diameter distances and Hubble parameter measurements from recent
BAO observations from BOSS DR12~\cite{bossdr12} compared with the corresponding quantities derived from PL18 
power spectra assuming a $\Lambda$CDM+$\Omega_K$ model.}
\label{Table1}
\end{table*}

As we can see from Table~\ref{Table2} the PL18 $\chi_{eff}^2$ best-fit is worse by $\Delta \chi^2\sim 16.9$ when the BAO data is included~\cite{planckwiki} under the assumption of curvature. This is a significantly larger $\Delta \chi^2$
than obtained for the  case of $\Lambda$CDM ($\Delta \chi^2\sim 6.15$). The BAO data set that we adopted consists of two 
independent measurements ( 6dFGS~\cite{6dFGS} and SDSS-MGS~\cite{mgs})
with relatively large error bars (i.e., with low statistical weight, see Figure~\ref{figure2}),  and $6$, correlated, measurements from 
BOSS DR12~\cite{bossdr12}. It is therefore not straightforward to determine the number of independent data points present
in the BAO dataset and to estimate the disagreement between the data sets from a simple $\chi^2$ analysis.
While several statistical methods have been proposed to quantify the discrepancy between two cosmological data sets~\cite{tool1,tool2,tool3,tool4}, here we check for consistency between two independent data sets $D_1$ and $D_2$ by evaluating the following quantity based on the DIC approach~\cite{Hildebrandt:2016iqg,joudaki16}:

\begin{equation}
{\mathcal{I}}(D_1, D_2) \equiv \exp\{{-{\mathcal{F}}(D_1, D_2)/2}\}, 
\label{eqn:logi1}
\end{equation}
\noindent where
\begin{equation}
{\mathcal{F}}(D_1, D_2) = {{{\rm{DIC}}(D_1 \cup D_2)} - {{\rm{DIC}}(D_1) - {{\rm{DIC}}(D_2)}}},
\label{eqn:logi2}
\end{equation}

\noindent where ${{\rm{DIC}}(D_1 \cup D_2)}$ is the DIC obtained from the combined analysis of the two data sets.

Following the Jeffrey's scale, the agreement/disagreement is considered 'substantial' if $|\log_{10} \mathcal{I}|>0.5$,
strong if $|\log_{10} \mathcal{I}|>1.0$ and 'decisive' if $|\log_{10} \mathcal{I}|>2.0$.
When $\log_{10} \mathcal{I}$ is positive then two data sets are in agreement while they  are in tension if this parameter is negative.
We show in Table~\ref{Table2} the values of $\log_{10} \mathcal{I}$ computed for the PL18 ($D_1$) and 
BAO ($D_2$) data sets in the case of $\Lambda$CDM and $\Lambda$CDM$+\Omega_K$. For the $\Lambda$CDM model, there is reasonable agreement between the data sets ($\log_{10} \mathcal{I}=0.2$), but evaluating models with curvature results in  substantial  disagreement $\log_{10} \mathcal{I}=-1.8$) between Planck and BAO data.

\begin{figure*}[!hbtp]
\begin{center}
\includegraphics[width=.7\textwidth,keepaspectratio]{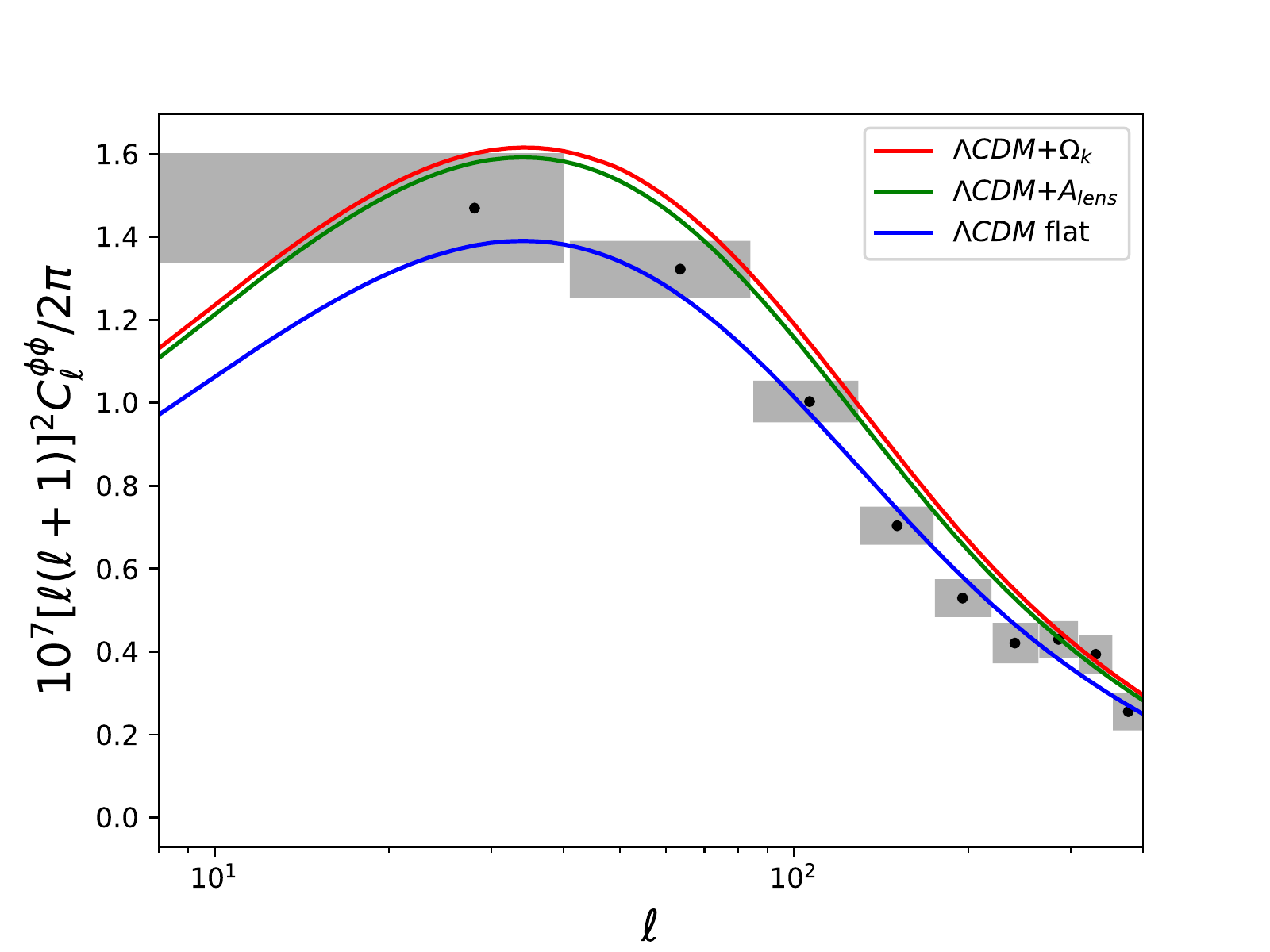}
\end{center}
\caption{{\bf Tension with CMB lensing}. The solid lines are the theoretical predictions for the best-fits CMB lensing power spectra from the PL18 angular spectra in case of  $\Lambda$CDM, $\Lambda$CDM+$\Omega_K$, and $\Lambda$CDM+$A_{lens}$ models, respectively. The
$\Lambda$CDM+$\Omega_K$ model has $\Omega_K=-0.0438$ while the flat $\Lambda$CDM+$A_{lens}$ model has $A_{lens}=1.191$.
The gray bands are the CMB lensing conservative experimental band powers extracted from the Planck 2018 trispectrum data.}
\label{figure2b}
\end{figure*}

A second tension is present between PL18 power spectra and the constraints on the lensing potential derived from the four-point function of Planck CMB maps\cite{plancklensing} (hereafter, CMB lensing). Indeed, as  discussed previously, the preference for $\Omega_K<0$ in PL18 is mostly due to the anomalous lensing amplitude at small angular scales~\cite{planck2018}. This greater lensing amplitude is however not seen in the CMB lensing data, which is consistent with flat $\Lambda$CDM. This can be seen in Figure~\ref{figure2b} where we compare the lensing-potential power spectra best fits from the PL18 power spectra, obtained under the assumptions of curvature or flatness, with the CMB lensing data~\cite{plancklensing}. The flat $\Lambda$CDM model is in reasonable agreement with CMB lensing, while the PL18 best fit $\Omega_K=-0.0438$ model predicts  too large a  lensing amplitude (with the exception of two data points). As we can see from Figure~\ref{figure2}, the PL18 power spectra best-fit closed model predicts a lensing-potential spectrum that is very similar to the best fit obtained under $\Lambda$CDM+$A_{lens}$ with $A_{lens}=1.191$\cite{planckwiki}.

A PL18+CMB lensing analysis yields $\Omega_K=0.011^{+0.013}_{-0.012}$ at $95 \%$ C.L., bringing a flat Universe back into agreement within  two standard deviations but  still also suggesting preference for a closed universe. It is however interesting to quantify the discordance between PL18 and CMB lensing. As we can see in Table~\ref{Table2}), the inclusion of CMB lensing to PL18 increases the best fit chi-square by $\Delta \chi^2 = 16.9$
 in the case of $\Lambda$CDM+$\Omega_K$ (while in the  case of $\Lambda$CDM model we have $\Delta \chi^2 = 8.9$). The CMB lensing data-set consists of $9$ 
 correlated data points. Even assuming these data points as independent, the increase in  $\chi^2$ when curvature is varied suggests there is  tension at the $95 \%$ C.L., while there is no significant tension in the case of flatness. 
Also in Table~\ref{Table2} we report the values of the $\mathcal{I}$ quantity. As we can see, we identify  substantial agreement between PL18 and CMB lensing in the case of a flat Universe ($\log_{10} \mathcal{I}=0.6$) that changes to "substantial discordance" ($\log_{10} \mathcal{I}=-0.55$) when curvature is allowed to vary .

\begin{table*}
\begin{center}\footnotesize
\begin{tabular}{|l|c|c|c|}
\hline
Additional dataset & $\Delta \chi^2_{eff}$ & $ \Delta N_{data}$& $log_{10}$ ${\mathcal{I}}$\\
\hline
{\it flat} $\Lambda$CDM &&&\\
\hline 
$+$BAO&$+6.15$&$8$&$0.2$\\
$+$ CMB Lensing&$+8.9$&$9$&$0.6$\\
\hline
$\Lambda$CDM+$\Omega_K$ &&&\\
\hline
$+$BAO&$+16.9$&$8$&$-1.8$\\
$+$ CMB Lensing&$+16.9$&$9$&$-0.84$\\
\hline
\end{tabular}
\end{center}
\caption{\bf Tensions between PL18 and BAO and CMB Lensing. In the second column we report the best fit $\Delta \chi^2_{eff}$ with respect to the PL18 dataset alone. In the third column the number $ \Delta N_{data}$ of (correlated) experimental data points from the additional data set. In the fourth column the value of $log_{10}{\mathcal{I}}$ that quantifies the tension (substantial if $<-0.5$, strong if $<-1$).  We can note  good agreement between the data sets in the  case of flat $\Lambda$CDM. 
On the contrary, statistically significant tensions arise when curvature is considered.}
\label{Table2}
\end{table*}

\begin{figure*}[!hbtp]
\begin{center}
\includegraphics[width=.57\textwidth,keepaspectratio]{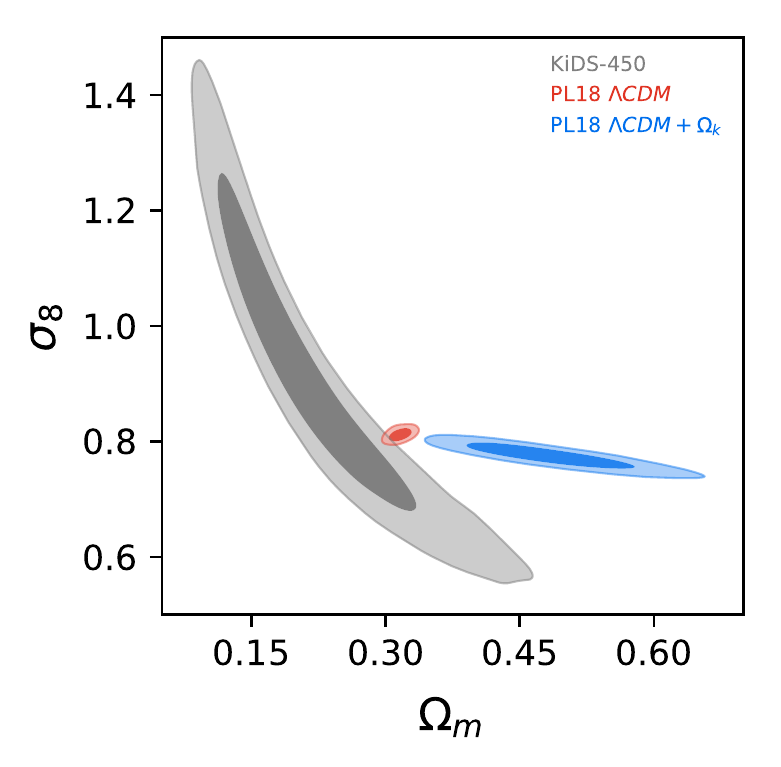}
\end{center}
\caption{{\bf Tension with cosmic shear measurements}. Discordance between PL18 and the KiDS-450 cosmic shear survey in the $\sigma_8$ vs $\Omega_m$ plane. The tension already present under flat $\Lambda$CDM (at about $2.3$ standard deviations) is increased to more than
$3.5$ standard deviations when curvature is incorporated into the analysis of PL18.}
\label{figure2c}
\end{figure*}

\begin{figure*}[!hbtp]
\begin{center}
\includegraphics[width=.45\textwidth,keepaspectratio]{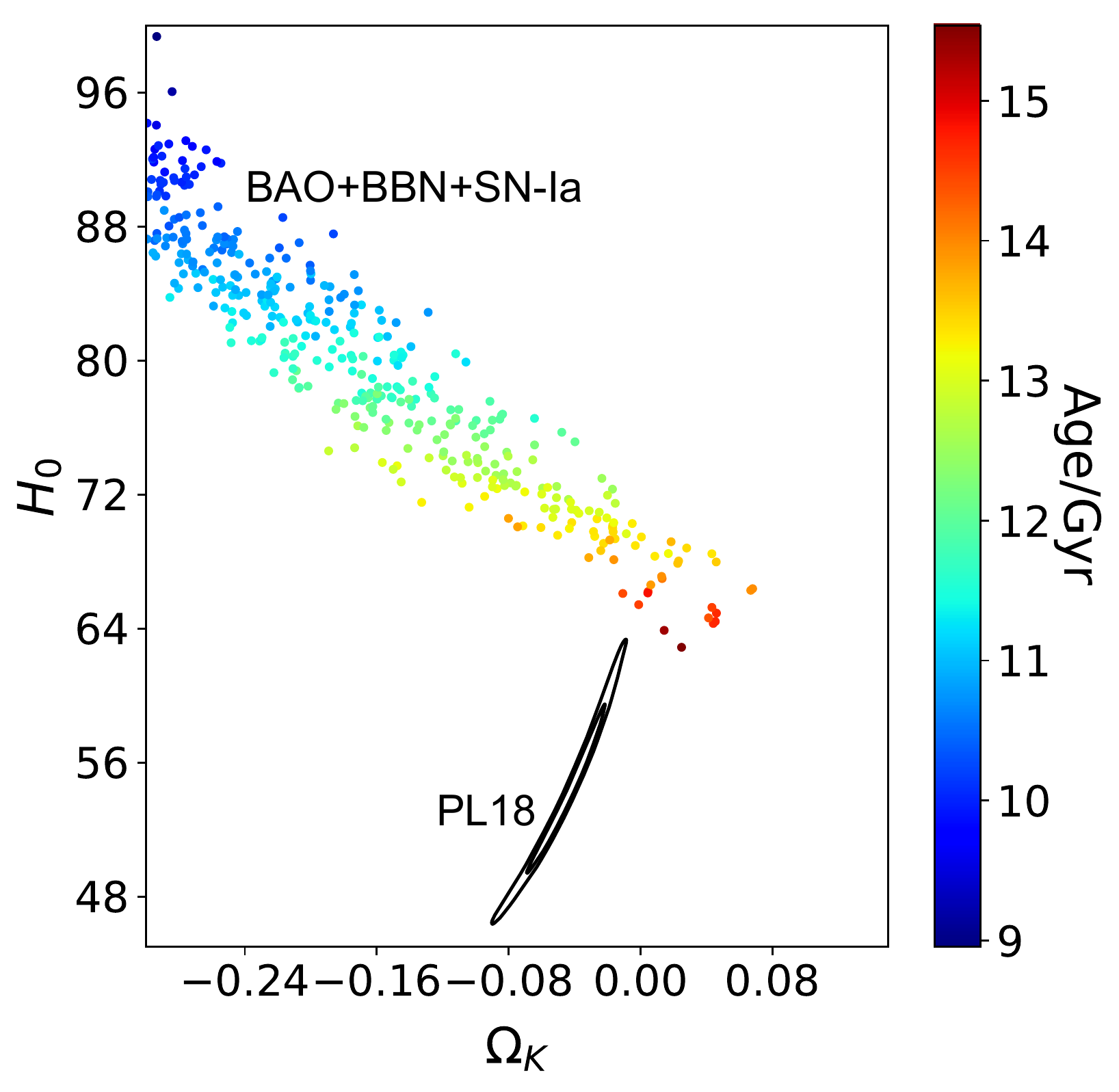}
\end{center}
\caption{{\bf Tension with combined data}. Contour plots at $68 \%$ and $95 \%$ C.L. from PL18 and BAO+SN-Ia+BBN data sets 
in the $H_0$ vs $\Omega_K$ plane and as a  function of the age of the universe.
}
\label{figure3}
\end{figure*}

In conclusion, if the assumption of a flat Universe is removed and curvature is permitted, as preferred by the PL18 power spectra, we find  strong disagreement between Planck and BAO, and  substantial disagreement between Planck and CMB lensing.

It is interesting to investigate if astrophysical measurements that are already in tension with Planck under the assumption of a flat Universe are still in disagreement when curvature is considered.  In a $\Lambda$CDM+$\Omega_K$ model, PL18 power spectra provide the constraint: $H_0=54.4^{+3.3}_{-4.0}$ at $68 \%$ C.L~\cite{planckwiki}. This is now in tension at the level of $5.2$ standard deviations with respect to the conservative R18 constraint of $H_0=73.52 \pm 1.62$ at $68\%$ C.L. \cite{riess2018}. The inclusion of curvature, therefore, significantly increases (by $\sim 48 \%$) the tension between Planck and R18.

A similar increase in the tension is present with cosmic shear data from KiDS-450. In Figure~\ref{figure2c} we show the 
2D constraints in the $\sigma_8$ vs. $\Omega_m$ plane from KiDS-450 and PL18 power spectra under the assumption of curvature. For comparison, we also include the Planck constraint under flat $\Lambda$CDM (the KiDS-450 bound is just slightly different when flatness is assumed).
 As we can see, there is a significant shift in this plane for the PL18 constraint when moving from a flat to a closed Universe, that increases the discrepancy with KiDS-450. From  the PL18 power spectra,  we obtain $S_8=0.981 \pm 0.049$ at $68 \%$ C.L. a value that is now about $3.8$ standard deviations from the KiDS-450 result. 
Cosmic shear measurements have also recently been made by the Dark Energy Survey~\cite{des} (DES) and by the Subaru Hyper Suprime-Cam~\cite{subaru} (HSC). These measurements are reasonably consistent with the PL18 result in a flat Universe. However, assuming that the reported constraint on the $S_8$ parameter depends weakly on $\Omega_K$, we find that once curvature is allowed, the PL18-derived determination of $S8$ is discordant at more than $3.5$ standard deviations with DES and at more than $3$ standard deviations with HSC. In practice, when curvature is included, not only the significance of the tension with KiDS-450 increases, but PL18 is now also significantly discordant with recent cosmic shear surveys as DES and HSC.

Until now we have studied the compatibility of single data sets with PL18. However analyses are usually performed by combining multiple datasets. It is interesting therefore to address the compatibility of Planck with combined data sets.
In Figure~\ref{figure3} we show the confidence region at $95 \%$ C.L. from a BAO+SN-Ia+BBN dataset and the $68 \%$ and $95 \%$ confidence levels from the PL18 power spectra on the $\Omega_K$ vs $H_0$ plane. As we can see, there is strong tension between the Planck result and that from the combined BAO+SN-Ia+BBN analysis. In principle, each data set prefers a closed Universe, with the BAO+SN-Ia+BBN data set providing just an upper limit of $\Omega_K<-0.124$ at $68 \%$ C.L. However, while the Planck result is preferring a value for the Hubble constant of $H_0=54_{-4.0}^{+3.3}$ km/s/Mpc at $68\%$ C.L., we find that the BAO+SN-Ia+BBN dataset gives $H_0=79.6\pm6.8$ km/s/Mpc at $68 \%$ C.L., i.e. they are inconsistent at the level of $3.4$ standard deviations.
Moreover, the BAO+SN-Ia+BBN data prefers lower ages of the universe with $t_0=11.73_{-1.3}^{+0.92}$ Gyr at $68 \%$ C.L. that is in modest tension with the the recent age determinations of the stars 2MASS J18082002--5104378 B~\cite{oldstar} and  HD 140283, of $t_*=13.535\pm0.002$ Gyr 
and  $t_*=13.5\pm0.7$ Gyr~\cite{hd140283,jimenez}, respectively

\begin{figure*}[!hbtp]
\begin{center}
\includegraphics[width=.57\textwidth,keepaspectratio]{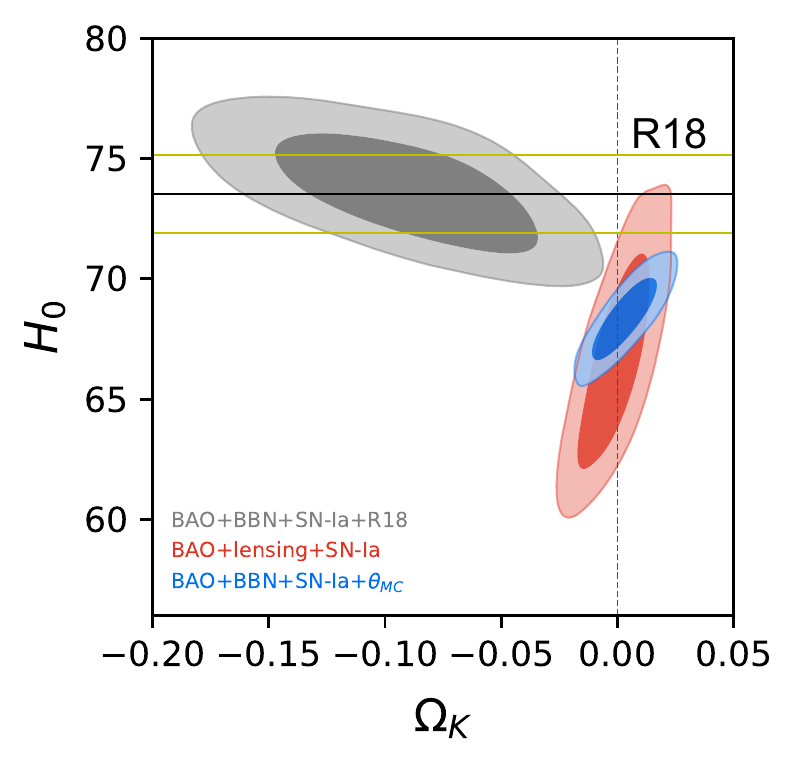}
\end{center}
\caption{{\bf Tensions in combined data}. Contour plots at $68 \%$ and $95 \%$ C.L. from the BAO+SN-Ia+BBN+$H_0$,
BAO+SN-Ia+BBN+$\theta_{MC}$,BAO+SN-Ia+CMB lensing datasets in the $\Omega_K$ vs $H_0$ plane.
Also shown in the figure are the constraints on the Hubble constant from R18~\cite{riess2018}.}
\label{figure4}
\end{figure*}

As we can see from Figure~\ref{figure3}, the probability contour plots from the BAO+SN-Ia+BBN analysis are rather broad. It is therefore interesting to include further observables to improve the constraints. We consider, separately, CMB lensing, the R18 determination of the Hubble constant and the observed angular size of the sound horizon at recombination $\theta_{MC}=1.04116\pm0.00033$ in a $\Lambda$CDM$+\Omega_k$ model from PL18. We show the results of this kind of analysis in Figure~\ref{figure4}. The inclusion of the $\theta_{MC}$ prior from Planck  shifts the constraints towards a flat $\Lambda$CDM with $\Omega_K=0.0016\pm0.0075$ at $68 \%$ C.L.. The inclusion of the CMB lensing dataset also significantly improves the constraints with $\Omega_K=0.00\pm0.01$ at $68 \%$ C.L..
Both BAO+SN-Ia+BBN+$\theta_{MC}$ and BAO+SN-Ia+CMB lensing combinations provide  evidence for a flat Universe, with good consistency between the data sets. We may argue that when deriving constraints under the assumption of a flat Universe, it would be more conservative to use these data combinations instead of PL18, since they are consistent with a flat $\Lambda$CDM and do not show significant internal tensions.  However, these data combinations still show a significant discordance with PL18 power spectra. Considering the parameter constraints derived from the BAO+SN-Ia+CMB lensing data set, we indeed find disagreement with PL18 at $2.4$ standard deviations in $\Omega_K$,  amounting to $2.7$ standard deviations in $H_0$, and $2.9$ standard deviations in $S_8$. 
When we consider the combination of BAO+SN-Ia+BBN+R18, we find $\Omega_K=-0.091\pm0.037$ at 68\% C.L., i.e. again providing an indication of a closed universe (see Figure~\ref{figure4}). Both data sets provide good best-fit chi-square values and it is impossible to discriminate one result over another from the statistical point of view. As we can see from Figure~\ref{figure4}, there is good agreement between the BAO+SN-Ia+CMB lensing and BAO+SN-Ia+BBN+$\theta_{MC}$ data sets while  both  are in in 
significant tension at the level of $2.5$ standard deviations with BAO+SN-Ia+BBN+R18.

In summary, the PL18 CMB power spectra provide a statistically significant indication for a closed universe. A closed universe solves the internal tensions present in the Planck data set on the value of the cosmological parameters derived at different angular scales. Positive curvature is also marginally suggested by the ages of the oldest stars (see e.g. \cite{oldstar,hd140283}) and, in a combined analysis with the $A_{lens}$ parameter, slightly favored by the low CMB quadrupole.

Apart from these arguments, none of the local cosmological observables currently favor a closed universe, and most of them are consequently  in significant discordance with PL18. BAO surveys disagree at more than $3$ standard deviations. CMB lensing is in tension at the level of $95 \%$ C.L. The R18 constraint on the Hubble constant is in tension with PL18 at more than five standard deviations, while cosmic shear data disagrees at more than $3$ standard deviations.

These inconsistencies between disparate observed properties of the Universe introduce a problem for modern cosmology: the flat 
$\Lambda$CDM, {\it de facto}, does not seem to any longer provide a good candidate for  concordance cosmology given the PL18 power spectra preference for a closed model. At the same time, a closed model is strongly disfavored by a large number of local observables.

Clearly, a possible solution to this problem would be to speculate about the presence of hitherto  undetected systematics 
in the PL18 release. However, the statistical significance for a closed Universe increases when moving from Planck 2015 to the PL18 release. We point out that  the WMAP satellite experiment~\cite{wmap9},  after $9$ years of observations, also produced the constraint $\Omega_K=-0.037^{+0.044}_{-0.042}$ at $68 \%$ C.L., fully compatible with the Planck result. 
Finally, we have shown that discordance is also present between the  R18 and the CMB lensing data sets once they are both combined with BAO and SN-Ia. In practice, there is currently no supporting evidence that could lead one  to believe that the observed inconsistencies are 
due to systematics in the PL18 data rather than  in the low redshift measurements. Moreover, local probes are expected to be more contaminated by astrophysical systematics and/or non-linearities with respect to CMB anisotropies.

If there are  indeed no systematics in the Planck data, then the currently observed discordances may indicate the need for new physics and call for drastic changes in the $\Lambda$CDM scenario (see e.g. \cite{newphys2,newphys4,newphys7,interactions1,interactions3,interactions5}).

A third possible way is to consider the PL18 constraint on $\Omega_K$ as a, now reasonably  unlikely, statistical fluctuation.
Fortunately, future measurements will fully confirm or falsify current tensions and the PL18 evidence for curvature~\cite{bull,bull2}. 
In the meantime, we argue that  the tensions with $\Lambda$CDM present in the PL18 release should not be discarded merely as a statistical fluctuation but must be seriously investigated, since at face value they  point towards a drastic rethinking of the current cosmological concordance model.

\begin{addendum}

\item[Correspondence] Correspondence and requests for materials should be addressed to A. Melchiorri \\ (email: alessandro.melchiorri@uniroma1.it).

\item[Acknowledgements]
EDV acknowledges support from the European Research Council in the form of a Consolidator Grant with number 681431.
AM thanks the University of Manchester and the Jodrell Bank Center for Astrophysics for hospitality. AM is supported by TAsP, iniziativa specifica INFN. We thank Michael Melchiorri for stimulating discussions. We thank Paolo De Bernardis, Dragan Huterer and Roberto Trotta for useful discussions.

\item[Note] This paper was submitted on the $10$th of March 2019 and accepted on the $5$th of September 2019.
During the review process a preprint by Will Handley (arXiv:1908.09139) appeared on August $24$ th, 2019 with similar conclusions.


\item[Author Information] 

\begin{affiliations}
\item Jodrell Bank Center for Astrophysics, School of Physics and Astronomy, University of Manchester, Oxford Road, Manchester, M13 9PL, UK
\item Physics Department and INFN, Universit\`a di Roma ``La Sapienza'', Ple Aldo Moro 2, 00185, Rome, Italy
\item Institut d'Astrophysique de Paris (UMR7095: CNRS \& UPMC- Sorbonne Universities), F-75014, Paris, France
\item AIM-Paris-Saclay, CEA/DSM/IRFU, CNRS, Univ. Paris VII, F-91191 Gif-sur-Yvette, France
\item Department of Physics and Astronomy, The Johns Hopkins University Homewood Campus, Baltimore, MD 21218, USA
\item BIPAC, Department of Physics, University of Oxford, Keble Road, Oxford OX1 3RH, UK
\end{affiliations}
 \item[Competing Interests] The authors declare that they have no
competing financial interests.
\end{addendum}

\clearpage

\end{document}